\documentclass[runningheads]{llncs}
\usepackage[T1]{fontenc}
\usepackage{graphicx}
\usepackage{xcolor}
\usepackage{longtable}
\usepackage{comment}
\usepackage{listings}

\begin{document}

\title{Large Language Models for Agent-Based Modelling: Current and possible uses across the modelling cycle}

\titlerunning{Large Language Models for Agent-Based Modelling}

\author{Lo\"is~Vanh\'ee\inst{1, 2}\orcidID{0000-0002-4147-4558} \and
Melania~Borit\inst{2}\orcidID{0000-0002-1305-8581} \and
Peer-Olaf~Siebers\inst{3}\orcidID{0000-0002-0603-5904} \and
Roger~Cremades\inst{4}\orcidID{0000-0002-4514-2462} \and
Christopher~Frantz\inst{5}\orcidID{0000-0002-6105-8738} \and
Önder~Gürcan\inst{6}\orcidID{0000-0001-6982-5658} \and
František~Kalvas\inst{7}\orcidID{0000-0002-9495-1602} \and Denisa~Reshef~Kera\inst{8}\orcidID{0000-0001-7242-9605} \and
Vivek~Nallur,\inst{10}\orcidID{0000-0003-0447-4150} \and
Kavin~Narasimhan\inst{11}\orcidID{0000-0002-5337-6736} \and
Martin~Neumann\inst{12}
}

\institute{Ume\aa{} Universitet; Ume\aa{}, Sweden \\\email{lois.vanhee@umu.se} 
\and UiT The Arctic University of Norway, Troms\o, Norway \\\email{melania.borit@uit.no}\\
\and University of Nottingham; Nottingham, UK\\\email{Peer-Olaf.Siebers@nottingham.ac.uk}\\
\and University of Leeds, Leeds, UK \\\email{ecosafor@gmail.com}\\
\and Norwegian University of Science and Technology, Gjøvik, Norway \\\email{christopher.frantz@ntnu.no}\\
\and NORCE Norwegian Research Center AS, Kristiansand, Norway \\\email{ongu@norceresearch.no}\\
\and University of West Bohemia, Plzeň, Czech Republic \\\email{kalvas@rek.zcu.cz}\\
\and Bar Ilan University, Ramat Gan, Israel \\\email{denisa.kera@gmail.com}\\
\and University College Dublin, Dublin, Ireland \\\email{vivek.nallur@ucd.ie}\\
\and University of Warwick, Coventry, UK \\\email{Kavin.Narasimhan@warwick.ac.uk}\\
\and University of Southern Denmark, Odense, Denmark \\\email{martneum@freenet.de}
}

%
%
 \authorrunning{L. Vanh\'ee et al.}
%
%
\maketitle              
\begin{abstract}
The emergence of Large Language Models (LLMs) with increasingly sophisticated natural language understanding and generative capabilities has sparked interest in the Agent-based Modelling (ABM) community. With their ability to summarize, generate, analyze, categorize, transcribe and translate text, answer questions, propose explanations, sustain dialogue, extract information from unstructured text, and perform logical reasoning and problem-solving tasks, LLMs have a good potential to contribute to the modelling process. After reviewing the current use of LLMs in ABM, this study reflects on the opportunities and challenges of the potential use of LLMs in ABM. It does so by following the modelling cycle, from problem formulation to documentation and communication of model results, and holding a critical stance. 

\end{abstract}
\section{Introduction}

 
The intersection of Agent-based Modelling (ABM) and Large Language Models (LLMs) presents a novel ground for advancing computational social science. ABM provides a bottom-up framework for simulating complex social phenomena through the behaviors and interactions of individual agents in a given environment \cite{bonabeau2002agent}. LLMs have powerful capabilities for engaging with text: summarization, translation, question answering, generation of human-like text, and conversational interaction \cite{orru2023human}. However, despite the proliferation of studies that place themselves in one way or another at this intersection, there is no structured approach that critically explores why, where, and how agent-based models and LLMs could be used together. This study addresses this gap by answering two research questions: \textbf{1.} \textit{Where and how are LLMs used in the ABM cycle?} \textbf{2.} \textit{Where and how could LLMs be used in the ABM cycle?} 

\vspace {2mm}


\vspace {-2mm}

\section{Previous relevant studies}

For the purpose of this work, several relevant previous studies were identified, some of them as pre-prints, indicating the fast-moving nature of the field. Hypothetical architectures and methods to systematically develop LLM-augmented social simulations are explored, and potential research directions in this field are discussed in \cite{gurcan2024llm}. 
The integration of LLMs in ABM in three ways (aiding in coding, generating synthetic data, and interpreting results) is discussed in \cite{polhill2025ethical}. 
Concrete advice for integrating LLMs in Agent-based Social Simulation model design is given in \cite{siebers2024exploring} and \cite{frydenlund2024modeler}, while four modelling and simulation tasks are covered in \cite{giabbanelli2023gpt}.
A significant attention is dedicated to the specific case of \textit{LLM-based agents}, i.e., using LLMs as a means to produce agent decisions (as a replacement for the classic software code) \cite{gao2024large}.
An analysis of the same kind of models can be found in \cite{larooij2025large}, this time focused on the ability of such models to address specific criticism towards conventional agent-based models in general: lack of realism, computational complexity, and challenges of calibrating and validating against empirical data.
How LLMs can be used for simulating human behavior in specific types of deliberation fields such as network science, evolutionary game theory, social dynamics, and epidemic modeling is discussed in \cite{lu2024llms}.
In addition, a demonstration of how to use LLMs when conducting simulation modeling research is given in \cite{akhavan2024generative}.

Although valuable for the community, these studies have several limitations. They are too narrow (focused on only one ability of LLMs, e.g., reproducing human-like behavior; focused only on one aspect of the agent-based model, e.g., agents and their behavior; or focused only on one step in the ABM process, e.g., model design); they do not follow a straightforward structure (neither in general nor for ABM); they follow a straightforward structure that is not specifically tailored for ABM, but for simulation modeling more broadly; or the structure they use does not cover the entire ABM cycle.
Our study aims to address these limitations by using the structure of the ABM cycle and exploring all the phases of this cycle and by accounting for the variety of LLMs abilities: summarize, generate, analyze, encode and categorize, transcribe and translate text, answer questions, propose explanations, sustain dialogue, extract information from unstructured text, and perform logical reasoning and problem-solving tasks.

\section{Conceptual background}

\subsection{Social simulation and Agent-based Modelling}

\textit{Social simulation} is a methodology for investigating complex social phenomena through the construction and analysis of artificial societies \cite{axelrod1997advancing}.
A common approach used within social simulation is \textit{Agent-based Modelling }(ABM), which involves modelling individual entities (agents) that represent social actors such as individuals, groups, or organizations. These agents operate within a defined environment and interact according to specified behavioral rules. ABM usually emphasizes bottom-up dynamics whereby macro-level patterns emerge from micro-level interactions of heterogeneous agents. Social simulation using ABM supports theory formulation \& testing, scenario analysis, and policy exploration, particularly in domains where empirical experimentation is difficult or unethical. 

\begin{figure}[t]\vspace{-1cm}
    \centering
    \includegraphics[width=0.9\linewidth]{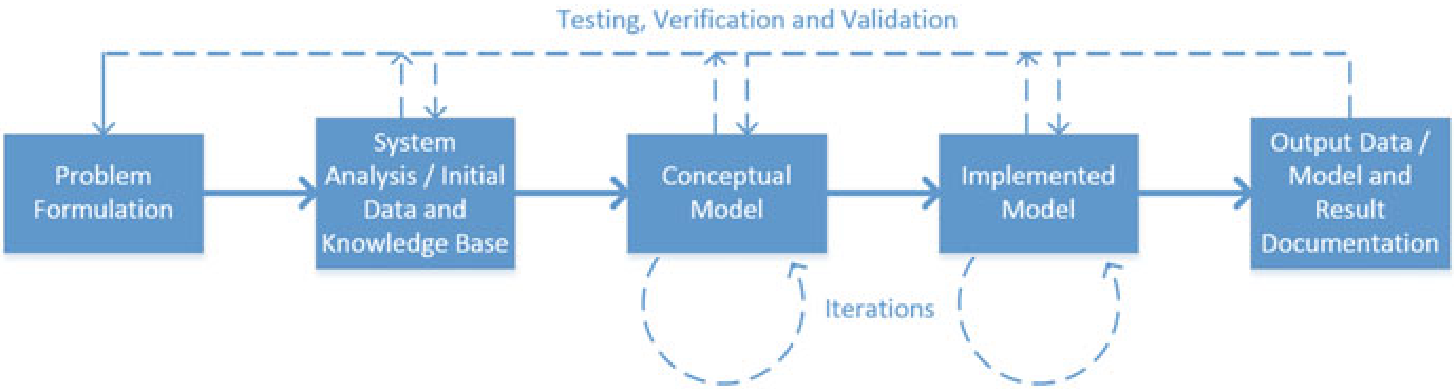}
    \vspace{-.2cm}
    \caption{Social simulation study life cycle. From \cite{siebers2017software}}
    \label{fig:life-cycle}
    \vspace{-.6cm}
\end{figure}

ABM is a multi-phase process involving abstracting from the world / phenomenon to the target system, understanding this system, abstracting concepts, modeling relationships, and creating simulation outputs. 
A generic \textbf{\textit{ABM cycle}} is presented in Figure \ref{fig:life-cycle}. Each phase consists of several iterative steps and sub-steps. This iterative process applies to the main phases (refinement through multiple cycles) and also within each phase by continuous improvement in feedback loops. 
For a detailed description of the phases of this cycle, see \cite{siebers2017software}
and \cite{nikolic2011method}.
The descriptions of the modelling phases from Section 5.2 are based on these two studies. When analysing these phases, one must keep in mind that in each of these several ethics-related questions can be asked \cite{anzola2022ethics} (e.g., Who brings input in the stage of formulating the research questions? Are assumptions communicated clearly and honestly? Was data collected in an ethical way?).

\vspace{-0.7mm}

\subsection{Large Language Models}

\textbf{\textit{Large Language Models (LLMs)}} are advanced neural networks pre-trained on vast textual datasets and used to process and generate human-like language \cite{patil2024review}
. They typically fall into two categories: generative models (e.g., GPT) and non-generative models (e.g., BERT) \cite{yang2024harnessing} 
(for a detailed explanation of how LLMs are built, their characteristics, contributions, and limitations, see \cite{minaee2025largelanguagemodelssurvey}
). Generative models predict the next token in a sequence, enabling them to produce coherent, contextually appropriate text. In contrast, non-generative models are commonly used for quantifying text into mathematical vectors. These vectors have the property of being geometrically close when the original texts from which they are derived are semantically close, making them well-suited for tasks like classification and retrieval of similar information. Although both capture deep linguistic patterns, only generative LLMs are designed for open-ended text generation and dialogue.

There are several main usage areas for LLMs \cite{yang2024harnessing}, including 
: 1) \textit{natural language understanding} (LLMs exhibit strong generalization capabilities); 2) \textit{text and code generation} (LLMs can produce coherent, contextually appropriate, and high-quality text and code, supporting applications such as content creation, programming assistance, and automated writing); 3) \textit{knowledge-intensive tasks} (LLMs can encode extensive and possibly implicit domain-specific knowledge, enabling them to perform tasks that require broad world knowledge or specialized expertise); 4) \textit{translations and transcriptions}; and 5) \textit{reasoning and problem solving} (LLMs can support complex tasks through pattern-based inference, analogical reasoning, and step-by-step problem decomposition). 
The use of LLMs has been associated with several \textit{ethical concerns} \cite{jiao2024navigating}, including 
bias  and  fairness,  privacy  and  data  security, misinformation  and  disinformation,  transparency  and  accountability,  intellectual  property  and plagiarism, access and inequality.

\vspace{-.2cm}
\subsection{Social Simulation through Agent-based Modelling and Large Language Models as part of Artificial Intelligence}
\vspace{-.1cm}

Before engaging in a detailed discussion regarding the potential role of LLMs in the various stages of the ABM cycle, with ABM acknowledged to be a common approach used within social simulation, it is important to situate these concepts within the broader Artificial Intelligence paradigm. 

Generative Artificial Intelligence (AI) represents the latest wave in the evolution of AI. In contrast, the 1990s saw the rise of ABM, often referred to as distributed AI, as a cutting-edge approach to building intelligent systems capable of exhibiting human-like behavior.
Broadly speaking, the field of AI has long been divided between two dominant paradigms: the symbolic and the data-driven (non-symbolic). While the symbolic tradition emphasizes logically encoded knowledge and rule-based reasoning, the data-driven approach prioritizes data-centric, statistical learning.
LLMs are emblematic of the data-driven paradigm, whereas ABM is more closely aligned with symbolic AI. 
A co-citation analysis by \cite{toosi2021brief} 
underscores the limited interaction between these two communities. Despite this separation, in this study, we ask whether these paradigms might now inform and enrich one another. Specifically, we explore how integrating LLMs into the ABM cycle could open up new opportunities, while also critically examining the risks and limitations such integration may entail.

Symbolic AI, considered the first generation of AI systems, aimed to model the world through predefined logical structures and algorithmic rules. Although this approach dominated early AI research, it was gradually overshadowed by data-driven approaches enabled by large-scale data availability and increased computational power.
Unlike symbolic AI, which formalizes reasoning through explicitly defined logic, data-driven AI identifies patterns in observations without encoding rules for each scenario \cite{khatami2025ai}. This distinction is rather important. Symbolic AI can theoretically exhaust all logical possibilities within its rule set, irrespective of empirical likelihood. In contrast, data-driven (or non-symbolic) AI is constrained by the scope and quality of its training data. Thus, while symbolic AI requires exhaustive logical modeling, data-driven systems depend on empirical coverage of real-world variation.
LLMs, as a subclass of data-driven AI, share these constraints. If training data lacks coverage or is of poor quality, models may produce inaccurate or “hallucinated” outputs. While retraining is often impractical, targeted supplementation with domain-specific data can help mitigate such issues. LLMs do not perform true forecasting but generate statistically likely continuations based on learned patterns.

\vspace{-4.5mm}

This distinction between symbolic and data-driven AI is important for several reasons. 
\textit{Social simulation requires agents to have decision-making mechanisms \cite{sun2018cognitive}
}. Symbolic AI fits this requirement, as the system follows predefined logic or heuristics. However, data-driven AI such as LLMs, while having the ability to emulate human-like responses, dialogue, and decisions based on linguistic patterns, it lacks consistent internal logic or goals. Thus, even if it can generate plausible behavior in rich environments, it struggles with coherent and consistent human-like reasoning. 
\textit{Social simulation emphasizes explicit model dynamics and model transparency \cite{carley2002simulating}
}. Symbolic AI allows for clear inspection of system's logic. In contrast, LLMs are black-box systems, making it difficult to trace the logic of why and how certain decisions were taken, as system action is shaped by training data distribution.
\textit{Social simulation often explores emergent or counterfactual scenarios \cite{gilbert2006emergence}
.} Symbolic AI enables precise specification of agent behavior, which makes it possible to  explore systematically how changes in assumptions or initial conditions influence outcomes. In contrast, being trained on large corpora of data, LLMs may generalize well to common-sense or culturally normative behavior, but are most likely to hallucinate or fail in niche domains, specific contexts, and emergent phenomena not captured by training data.
\textit{Social simulation is interested in capturing implicit social norms and tacit knowledge \cite{edmonds2013agent}
.} Such aspects might be more easy captured by non-symbolic AI than symbolic systems.
\textit{Social simulation needs flexibility in loosely-structured cases \cite{edmonds2013agent}
.} In such scenarios or exploratory simulations, where behavior is not easily defined by strict rules, non-symbolic AI offers a flexible way to generate plausible agent behavior without fully specifying all conditions (something that a symbolic system would need). 


\section{Method}

A Rapid Literature Review (RLR) was used to answer the first research question, i.e., "Where and how are LLMs used in the ABM cycle?". A RLR is a streamlined form of a systematic review that balances rigor with efficiency to provide timely evidence. The RLR phases described in \cite{Bouck2021} were used to identify, select, critically appraise, and synthesize documents meeting pre-specified eligibility criteria, as described below. 
Data for the RLR was last collected in March 2025, using one single database, i.e., Scopus, which is an abstract and citation database of peer-reviewed scientific literature produced by Elsevier. The search query can be found in the annex.

The papers retrieved by the search were excluded if: they were not in English; they were reviews, position papers, and conference summaries; they did not include a description of an implemented use of LLMs in connection with ABM; or if no agent-based model was introduced in the paper.

Two of the co-authors screened titles and abstracts for inclusion/exclusion and performed the paralleled strict screening of full-text articles that passed the broad screening step. They performed data extraction (which followed a coding scheme developed based on the research questions) and data synthesis. The coding scheme included codes related to: meta data (publication year and domain of publication venue, as categorized by Scopus), type of study (focused on methodology; focused on application); where in the ABM cycle (cf. \cite{siebers2017software}
) were LLMs used; whether ethical aspects of using LLMs were discussed. In order to ensure inter-coder reliability, the two coders checked each other's coding and discussed it extensively.

Group discussions were used to answer the second research question, i.e., question "Where and how could LLMs be used in the ABM cycle?". Starting with June 2024, the co-authors had regular roughly monthly meetings organized by one of the co-authors (who recruited interested people using various email lists, e.g., SimSoc). The co-authors have very different backgrounds: computer science, operation research, sociology, philosophy, design, natural resource management, and climate change. Their experience with LLMs, ABM, or LLMs and ABM varies from very limited to advanced.

\section{Results and discussion}

\subsection{Where and how are Large Language Models used in the Agent-based Modelling cycle?}

The search query identified 40 items, out of which 22 were kept for in depth coding. Two of these are published in 2023, 17 in 2024, and three in 2025. Almost 80\% of the papers (n=17) are published in computer science, three in engineering \cite{lovaco2024llmfirefighting} \cite{ilagan2024customer} \cite{zhang2025llmaidsim}, and one in each mathematics \cite{ferraro2024agentbased} and social sciences \cite{ghaffarzadegan2024generative}. This trend is to be expected considering the technical nature of LLMs. Five papers are published in a journal, with a vast majority (almost 80\% ) being published in conference proceedings. Again, this trend is to be expected considering the dynamism of the developments in the LLM domain. The applicative domains vary from general to microeconomic decisions, information propagation, and norm diffusion. The vast majority of the papers are methodological. 
Only two papers include ethical considerations \cite{ilagan2024customer} \cite{wu2024teamup}.

The most common use (n=20, 91\%) involves implementation, particularly focusing on agents powered by LLMs for reasoning, deliberation /decision-making, or communication.
Code generation (in NetLogo) is the main focus in one  study \cite{martinez2024gptnetlogo}. 
Interpretation of model results is the main focus in one study using LLMs to generate narrative explanations that communicate agents’ perspectives on key life events \cite{lynch2023structured}. 

As such, it can be concluded that the inherent potential of LLMs has not yet been fully leveraged throughout the ABM cycle.


\vspace{-1mm}

\subsection{Where and how could Large Language Models be used in the Agent-based Modelling cycle? }
  

\subsubsection{\textbf{Problem formulation:}}

\textbf{\textit{Why?}} The problem must be well-described, as research questions are formulated based on that. Especially when the modeler is not sufficiently acquainted with disciplinary terminology (e.g., multi-, inter-, and transdisciplinary studies); comprehension and synthesis of large bodies of knowledge, as well as interaction with stakeholders (e.g., participatory research) are challenges in this phase. 
\textbf{\textit{How?}} LLMs can possibly assist as follows: 
(1) \textit{conducting literature reviews} by summarizing research findings across disciplines; 
(2) \textit{supporting communication with stakeholders} through natural language dialogues to gather goals and constraints; 
(3) \textit{generating and evaluating hypothetical scenarios} to help refine problem definitions, objectives, and research questions for the simulation; 
and (4) \textit{translate text} (e.g., literature, stakeholders input) into the working language of the modeler. 
\textbf{\textit{Potential pitfalls?}} Potential risks include overreliance on LLM-generated content, which may introduce biases from the training data (e.g., implicitly attributing certain characteristics depending on the gender or ethnicity of the agent), hallucinate facts, reduce the breadth of possible models and research activities to pre-established norms. Additionally, the outputs might lack domain-specific nuance or fail to capture contextual priorities if used uncritically.
\textbf{\textit{How to mitigate?}} Modelers should critically evaluate LLM outputs, combine them with expert knowledge, and validate findings through consultation with stakeholders and domain experts. 

\vspace{1mm}

\subsubsection{\textbf{System analysis:}}
\textbf{\textit{Why?}} This phase in the modelling cycle is critical to understand what is to be modelled: key entities, interactions, and emergent dynamics within the system to be modeled.
It guides the abstraction process and it helps identifying the internal structure of the system under analysis, as well as its boundaries, agent behaviors, environmental factors, and relationships. As the components of the system are identified, they also need to be arranged into a structure that is manageable and understandable by the modeller (and other participants in the modelling process). 
When building this structure, concepts are harmonized. 
Sources of information and data are looked for in this phase.
However, challenges arise when dealing with complex or poorly structured domains (e.g., domains where information is ambiguous, incomplete, or lacks a clear theoretical or formal framework), especially when system components span diverse fields or when the underlying causal mechanisms are not well understood. 
Additionally, modelers may struggle to extract or synthesize relevant mechanisms from extensive and heterogeneous literature or stakeholder narratives.
\textbf{\textit{How?}} LLMs can possibly assist as follows: 
(1) \textit{extracting mechanisms} (parsing large volumes of domain-specific literature to identify relevant theories and summarize relevant processes, feedback loops, mechanisms, and causal relationships) and \textit{generating candidate system structures}, including agent typologies, interaction rules, and environmental features (based on descriptive text, such as field notes or stakeholder inputs);
(2) \textit{conceptual mapping} (generating or supporting the development of influence diagrams or interaction schematics based on textual descriptions); 
(3) \textit{hypothesis generation} (proposing plausible agent rules or system mechanisms, including alternative explanations for observed behaviors or empirical patterns); 
(4) \textit{cross-domain integration} (facilitating connections between theoretical frameworks or empirical findings from different disciplines to enrich system representations);
(5) \textit{providing contextual argumentation} (based on previous models or on primary data, by offering narratives explaining agent behaviors in given scenarios or by imitating the view points of various stakeholders).
\textbf{\textit{Potential pitfalls?}} There are risks of misinterpreting or oversimplifying complex dynamics when relying on LLMs. They might suggest mechanisms that sound plausible but lack empirical support or misrepresent causal directionality. Additionally, LLMs may under-represent minority perspectives or context-dependent behaviors, or may fail to identify subtle feedbacks, emergent properties, or power asymmetries.
\textbf{\textit{How to mitigate?}} To reduce these risks, LLM outputs should be treated as provisional aids, not definitive sources. Key mechanisms suggested by LLMs should be triangulated with domain-specific literature, human stakeholder input, and expert reviews. Maintaining transparent documentation of assumptions and justifications for included system elements also helps ensure scientific rigor.

\vspace{-2mm}

\subsubsection{\textbf{Conceptualization:}} 
\textbf{\textit{Why?}} The conceptual model defines the abstract representation of the system to be simulated, including agents, their attributes, behaviors, and interactions, in a computationally feasible way. It acts as a bridge between the abstract understanding of the system to be modelled and computational implementation. This phase requires organizing diverse information sources into a coherent structure, which often demands both disciplinary insight and creative synthesis. For modelers unfamiliar with certain domain-specific logics or frameworks, this translation from system definition to simulation-ready abstraction can be particularly challenging.
\textbf{\textit{How?}} 
LLMs can possibly assist as follows: 
(1) \textit{agent and environment specification} (translating textual system descriptions into structured outlines of agents, their goals, decision-making processes, and the properties of the environment they inhabit);
(2) \textit{rule formulation} (suggesting candidate behavioral rules for agent actions based on theoretical inputs, empirical cases, or stakeholder narratives);
(3) \textit{model articulation} (helping to draft readable and structured descriptions of the model using established templates such as inherent in the ODD protocol \cite{grimm2010odd});
(4) \textit{feedback loop formalization} (identifying and refining hypothesized causal pathways and feedback structures embedded in agent interactions and environmental dynamics);
(5) \textit{consistency checking} (evaluating whether the elements and assumptions of the model are logically coherent and aligned with the intended theoretical framing).
LLMs can assist in identifying central and peripheral concepts, as well as the entities/actions/properties that link them together. Through parsing of large volumes of multi-modal data, LLMs may be able to connect concepts across distributed knowledge sources, which may be hitherto unknown, or rarely juxtaposed. The summarization ability of LLMs could also lead to a reduction in time-to-understanding, even if this is at a superficial level, initially, for experts from differing domains. Considering the iterative nature of conceptualization (from a process perspective) and the large, rich content (from a quality perspective), issues such as document inconsistency, lack of comprehensiveness (missing details and uncovered aspects), and weak connections between phenomena can arise, areas where LLMs can be highly valuable in the standardization process through rule checking, document correction, and factor linkage.
\textbf{\textit{Potential pitfalls?}} LLM-generated structures might reflect patterns seen in training data rather than the unique logic of the specific system under study. There is also a risk of incorporating implicit biases or oversimplified mechanisms that do not align with empirical realities or stakeholder perspectives. Moreover, LLMs may struggle with internal coherence or inadvertently exclude critical contextual factors.
Furthermore, the suggestions made by LLMs may be grounded on expected factors and may introduce subtle logical inconsistencies, in particular if the LLM is not tasked with explaining its decisions.
This lack of transparency directly conflicts with modelling emphasis on inspectable and verifiable system identification, as promoted by standards like the ODD protocol \cite{grimm2010odd}. 
\textbf{\textit{How to mitigate?}} Modelers should iteratively refine LLM-supported conceptual models by comparison with established theories, empirical data, and stakeholder feedback. Extensive justification for the various suggestions should be demanded from the LLM and scrutinized by the modeller, for ensuring the defensibility of the conceptualization.
Engaging domain experts to review generated elements can help ensure relevance and accuracy. All conceptual assumptions should be explicitly documented to maintain transparency and support reproducibility.

\vspace{-2mm}

\subsubsection{\textbf{Implementation:}}
\textbf{\textit{Why?}} 
The implementation of an ABM is arguably the least creative step, and also, possibly, the most susceptible to errors due to translation errors from the modeller's view into the computer's view. For large domains with complex inter-relationships, the amount of programming involved can be non-trivial. Small mistakes in programming could lead to errors in model-building, or worse, errors in simulation that may not be detectable. Challenges may arise in maintaining traceability from model design to code, especially in large or modular systems, and in ensuring that implementation choices do not introduce unintended artifacts or biases.
\textbf{\textit{How?}} 
LLMs can possibly assist as follows: 
(1) \textit{code generation} (translating structured model descriptions—such as ODD or other structured design documents—into executable code in platforms like NetLogo or Repast, possibly using libraries that the modeller is unfamiliar with or unaware of; the speed achieved in implementation could be orders of magnitude greater than achievable through human efforts, and this could lead to reduced iteration times, and thereby increased number of iterations);
(2) \textit{template set up} (producing templates or skeleton code for agents, environments, schedulers, or user interfaces, reducing time spent on repetitive setup);
(3) \textit{code explanation and code cleaning} (clarifying or restructuring existing model code to improve readability, modularity, or performance);
(4) \textit{use of LLM-powered agents* (see next paragraph for more details about LLM-powered agents)} (embedding LLMs as cognitive components within agents to simulate natural language reasoning, narrative behavior, or adaptive decision-making in complex or poorly defined environments);
(5) \textit{debugging support} (helping diagnose and resolve common programming errors or inconsistencies between intended and observed behaviors);
(6) \textit{documentation and commenting} (generating inline code comments or technical documentation that maintains coherence with the conceptual model);
(7) \textit{train novice programmers}.
\textbf{\textit{Potential pitfalls?}} 
With a large code-base, it would be difficult to ensure that the agents' behavior is what the modeller intended. Overreliance on LLMs might reduce understanding of the model. Lack of intimate familiarity with the code-base could lead to a downward slope of ability/expertise to modify the model in any meaningful manner. Moreover, LLM-generated code may contain subtle bugs, inefficient structures, or mismatches with the intended logic of the conceptual model.
\textbf{\textit{How to mitigate?}} 
Start with simple prototype and add complexity iteratively. A regular code walkthrough of every model iteration would ensure that the modellers stay in touch with the generated code-base, and are able to follow the program's logic. Moreover, all LLM-generated code should be treated as a first draft. Such code must be reviewed and tested thoroughly. Use mainstream libraries. Add an extra column in Rigour And Transparency – Reporting Standard (RAT-RS) \cite{achter2022ratrs} to document the use of LLMs.

\vspace{-2mm}

\subsubsection{\textit{\textbf{\textit{*The case of LLM-powered agents:}}}}

\textit{\textbf{Why:}} Human actors have strong inference capacities based on the level of natural language. For actors embedded in human cultural settings, the negotiation of meaning takes place in a triadic relation between objects, symbols, and human practices. Such context-sensitive language level communication is only barely captured by agents in traditional agent-based models. For instance, in opinion dynamics models, the influence of agents on each other is mostly represented by a shift of numerical values depending on certain influence probabilities that might be implemented in different rule sets depending on whether the model follows, for example, the bounded confidence approach, or may, for example, include repulsive forces between agents with extremely different opinions (see e.g., \cite{flache2017models}). However, nuances of expressions that might be contextually very relevant cultural settings cannot be captured by a purely numerical approach to communication. 
\textbf{\textit{How:}} Generating natural language texts is the prime strength of LLMs. Therefore it might be useful to integrate an LLM into an agent-based model or coupling an agent-based model with an LLM inference machine in which the agent-based model component sends a request to the LLM and the LLM returns a text in natural language. If agents exchange such messages and reason about and respond to received messages via an LLM component, communication in natural language can be integrated in an agent-based model.
\textbf{\textit{Potential pitfalls:}} It is not a priori evident that chats generated by LLMs factually represent a discourse that potentially could be found in the target system. In particular, as LLMs are based on training data, they might be less able to replicate a specific feature of intelligence distributed in human discourse: namely, the potential not just to reflect given knowledge but rather that the discursive dynamics might foster to find novel and innovative answers or problem solutions that have not been thought of before.
\textbf{\textit{How to mitigate:}} The iterative nature of the ABM cycle might be useful: a discourse artificially generated by an LLM can be presented to and discussed by relevant stakeholders in the field.

\vspace{-2mm}

\subsubsection{\textbf{Verification:}}
\textbf{\textit{Why?}} Verification ensures that the model behaves as intended and that its implementation is complete and aligns with the underlying design specifications - it addresses the question of building the model right \cite{robinson1997verification}. The model is checked against its conceptual design to see if all relevant entities and relationships from the conceptual model were correctly translated into the computational model.
\textbf{\textit{How?}} 
LLMs can possibly assist as follows: 
(1) \textit{code review assistance} (highlighting potential logic errors, missing edge case handling, or inconsistencies between code and documentation);
(2) \textit{trace interpretation} (analyzing simulation logs or stepwise outputs to detect anomalies or deviations from expected behavior; generating stories based on rules executed by agents that are afterwards checked for meaningfulness by the modeller);
(5) \textit{comparison tools} (automating comparisons between multiple simulation runs to detect instability, randomness artifacts, or implementation inconsistencies).
\textbf{\textit{Potential pitfalls?}} 
LLMs may overlook domain-specific logic errors or flag harmless patterns as problematic due to a lack of context. LLM-generated interpretations of simulation traces may reflect surface-level associations rather than deep causal understanding. Automatically generated agent “stories” may appear plausible, but could lack internal consistency or fail to reflect the actual rule execution logic. Automated comparisons may focus on superficial output differences.
\textbf{\textit{How to mitigate?}} 
Use LLM as a complement, not replacement of manual or peer-based code reviews. Review LLM-generated narratives alongside rule logs. Use multiple forms of comparison.

\vspace{-2mm}

\subsubsection{\textbf{Validation}} 
\textbf{\textit{Why?}} 
Validation assesses whether the model accurately represents the real-world system it is intended to simulate — it addresses the question of building the right model \cite{robinson1997verification}. It can involve comparison with empirical data, expert judgment, stakeholder feedback, theoretical expectations, or existing models.
\textbf{\textit{How?}} 
LLMs can possibly assist as follows: 
(1) \textit{empirical alignment} (helping identify relevant datasets or empirical patterns for comparison with model outputs and assisting in the generation of hypotheses about alignment or mismatch);
(2) \textit{narrative validation} (generating or analyzing scenario-based narratives that describe plausible sequences of agent behavior and comparing these with stakeholder accounts or case studies);
(3) \textit{expert emulation} (imitating expert reasoning or stakeholder perspectives to evaluate whether model outcomes are perceived as realistic or plausible);
(4) \textit{triangulation support} (synthesizing findings from multiple sources, e.g., literature, data, or stakeholder input, to inform decisions about the model’s validity).
\textbf{\textit{Potential pitfalls?}} 
LLMs may overgeneralize or misinterpret empirical findings, especially when data is sparse, noisy, or context-dependent. They may also suggest overly plausible interpretations that hide underlying model deficiencies. Narrative-based or qualitative validations supported by LLMs can be biased toward familiar or dominant frames, potentially missing outlier cases or minority viewpoints. 
\textbf{\textit{How to mitigate?}} 
Validation efforts should remain grounded in established best practices: comparing model outcomes with independent data and existing models, engaging with domain experts, and documenting the rationale behind validation choices. LLM contributions should be treated as exploratory tools. 

\vspace{-2mm}

\subsubsection{\textbf{Interpretation and communication of results:}}
\textbf{\textit{Why?}} 
This is the phase where insights are extracted from simulation output. Traditionally, agent-based models are developed to understand how macro-level social structures can be generated from the bottom-up by interacting agents. Typically, these structures are described by time series or other numerical results. However, for descriptive modelling purposes, it might also be useful to delve into micro-level details of the results. Investigating the meaningfulness of the simulation with the help of LLMs opens new perspectives for ethnographic, interpretive research questions for agent-based social simulation.
\textbf{\textit{How?}} 
LLMs can possibly assist as follows: 
(1) \textit{pattern discovery} (identifying trends, clusters, or tipping points in outputs);
(2) \textit{narrative synthesis} (generating structured or story-based explanations of model behavior over time, helping to communicate how outcomes arise from 
agent 
interactions and dynamics);
(3) \textit{stakeholder-tailored reporting} (adapting result summaries to different audiences, including policymakers or community groups; translating the summaries in different languages);
(4) \textit{sensitivity and robustness reflection} (helping describe the implications of parameter sensitivity and robustness checks in plain language);
(5) \textit{cross-scenario comparison} (summarizing differences and similarities across multiple simulation runs or policy scenarios to highlight key drivers and trade-offs).
\textbf{\textit{Potential pitfalls?}} 
LLMs may oversimplify or overinterpret outputs or present patterns where none exist.
They may introduce narrative bias. In comparative scenarios, they may highlight surface differences while missing deeper, systemic mechanisms. There is a risk that modelers rely too heavily on LLM summaries.
\textbf{\textit{How to mitigate?}} 
Results should be interpreted in conjunction with domain knowledge and statistical validation. Use LLM outputs as help to guide exploration, not as standalone conclusions. When generating narratives, cross-reference them with logged behaviors and rules to ensure internal consistency. Double-check with experts/stakeholders.

\vspace{-2mm}

\subsubsection{\textbf{Documentation:}}
\textbf{\textit{Why?}} 
Documentation ensures transparency, reproducibility, replicability, and transferability of the model. It records the assumptions, design decisions, data sources, and implementation details that shaped the model and its outputs. Though moften considered daunting by the modellers, in ABM this phase is very important because of the complexity of the modelling process.
\textbf{\textit{How?}} 
LLMs can possibly assist as follows: 
(1) \textit{automated report generation} (summarizing model structure, parameters, and outputs in clear language from code bases or ODD descriptions);
(2) \textit{commenting and inline explanations} (generating or refining explanatory comments within code files to improve readability and follow up);
(3) \textit{version summaries} (tracking and describing changes across model versions or iterations, including rationale for changes);
(4) \textit{audience-adapted documentation} (transforming technical content into stakeholder-facing formats, such as executive summaries or policy briefs; translate the summaries in different languages).
\textbf{\textit{Potential pitfalls?}} 
Too generic or templated text may fail to reflect the specific structure or intent of the model. There is also a risk of inconsistencies between code, conceptual design, and generated explanations.
\textbf{\textit{How to mitigate?}} 
Use LLMs to augment, not replace, human-authored documentation. Always review and edit generated text for accuracy, completeness, and clarity. Maintain documentation as a living artifact, updated in parallel with model development. Employ structured protocols  (e.g., ODD, RAT) to guide LLM-supported documentation efforts and ensure alignment with best practices. As with conventional documentation, version control systems and collaborative review processes can enhance this phase.


\vspace{-2.3mm}
\section{Conclusion}
\vspace{-2.1mm}

The methodological tensions highlighted above are not merely technical but also ontological. ABMs fundamentally seek to formalize systems through explicitly defined rules and entities, while LLMs operate through emergent patterns recognized within natural language. Reconciling these paradigms requires acknowledging that LLMs may not directly replace traditional processes within the ABM cycle. Instead, their value lies in their potential to enrich, provoke, or challenge existing system conceptualizations held by modellers and stakeholders. They become useful not primarily as suppliers of structured model input, but as tools that expand the interpretive lens brought to the modelling problem. They integrate disparate data and automate preliminary modeling tasks, extract implicit semantic relationships that illuminate system constraints, and support rapid generation of models subject to expert verification. Rather than directly verifying system variables, LLMs provoke rethinking and refinement of system models, offering new perspectives that may benefit modelers and stakeholders. By integrating LLMs cautiously and critically in the ABM cycle, checking their outputs rigorously, modellers may unlock new opportunities for more participatory and pluralistic forms of modeling. In this vision, the formal logic of ABM and the narrative potential of LLMs coexist not as competitors, but as complementary co-constructors of system understanding.

\section{Acknowledgments}

LV acknowledges the financial support of the project ``Anxiety-Sensitive Artificial Intelligence", funded by the Swedish Research Council (project number 2023-04505), Sweden. MB  acknowledges the financial support of the project ``FUTURES4Fish: Adaptive socio-technological solutions for Norwegian fisheries and aquaculture'', funded by the Research Council of Norway (project number 325814). FK acknowledges the institutional support of the University of West Bohemia. VN acknowledges that this publication has partially emanated from research conducted with the financial support of Irish Research Council via Grant COALESCE/2021/4, and the Research Ireland Centre for Research Training in Digitally-Enhanced Reality (d-real) under Grant No. 18/CRT/6224 and and Grant No. 18/CRT/6183'.  For the purpose of Open Access, VN has applied a CC BY public copyright licence to any Author Accepted Manuscript version arising from this submission.

\bibliographystyle{splncs04}
\bibliography{biblio}

\section*{Annexes}

\subsection*{Rapid Literature Review query}

\noindent
\begin{lstlisting}
(TITLE-ABS("chat gpt" OR "chatgpt" OR "large language model" OR "llm" OR "prompt engineering") OR 
SRCTITLE("chat gpt" OR "chatgpt" OR "large language model" OR "llm" OR "prompt engineering") OR
(AUTHKEY("chat gpt" OR "chatgpt" OR "large language model" OR "llm" OR "prompt engineering")
AND (TITLE-ABS("agent-based model*" OR "agent-based simulation" OR "agentbased model*" OR "agentbased simulation" OR "individual-based model*" OR "social simulation")
OR SRCTITLE("agent-based model*" OR "agent-based simulation" OR "agentbased model*" OR "agentbased simulation" OR "individual-based model*" OR "social simulation")
OR AUTHKEY("agent-based model*" OR "agent-based simulation" OR "agentbased model*" OR "agentbased simulation" OR "individual-based model*" OR "social simulation"))
\end{lstlisting}

\newpage

\subsection*{List of papers coded during the Rapid Literature Review}

\begin{longtable}{p{12cm}|l}
    \caption{List of papers coded during the Rapid Literature Review}        
    \label{table:all_papers}
  \\
    \textbf{Title} & \textbf{Reference}  \\\hline
        Analysis of LLM-Based Narrative Generation Using the Agent-Based Simulation & \cite{aoki2023llm}\\
A Structured Narrative Prompt for Prompting Narratives from Large Language Models: Sentiment Assessment of ChatGPT-Generated Narratives and Real Tweets & \cite{lynch2023structured}\\
Enhancing Gpt-3.5’s Proficiency In Netlogo Through Few-Shot Prompting And Retrieval-Augmented Generation & \cite{martinez2024gptnetlogo}\\
Spontaneous Emergence of Agent Individuality Through Social Interactions in Large Language Model-Based Communities &\cite{takata2024individuality}\\
Solution-oriented Agent-based Models Generation with Verifier-assisted Iterative In-context Learning & \cite{niu2024solution}\\
Simulating Opinion Dynamics with Networks of LLM-based Agents &\cite{chuang2024simulating}\\
Unveiling the Truth and Facilitating Change: Towards Agent-based Large-scale Social Movement Simulation &\cite{mou2024unveiling}\\
Large language model-driven simulations for system of systems analysis in firefighting aircraft conceptual design &\cite{lovaco2024llmfirefighting}\\
Learning Agent-based Modeling with LLM Companions: Experiences of Novices and Experts Using ChatGPT \& NetLogo Chat &\cite{chen2024learning}\\
Prototyping Slice of Life: Social Physics with Symbolically Grounded LLM-based Generative Dialogue & \cite{treanor2024slice}\\
Exploratory Customer Discovery Through Simulation Using ChatGPT and Prompt Engineering &\cite{ilagan2024customer}\\
An LLM-enhanced Agent-based Simulation Tool for Information Propagation &\cite{hu2024llmpropagation}\\
EconAgent: Large Language Model-Empowered Agents for Simulating Macroeconomic Activities &\cite{li2024econagent}\\
TraderTalk: An LLM Behavioural ABM applied to Simulating Human Bilateral Trading Interactions &\cite{vidler2024tradertalk}\\
Paradise: An Experiment Extending the Ensemble Social Physics Engine with Language Models &\cite{kelly2024paradise}\\
Is this the real life? Is this just fantasy? The Misleading Success of Simulating Social Interactions With LLMs &\cite{zhou2024misleading}\\
Enhancing spatially-disaggregated simulations with large language models &\cite{zaslavsky2024enhancing}\\
Shall We Team Up: Exploring Spontaneous Cooperation of Competing LLM Agents & \cite{wu2024teamup}\\
Generative agent-based modeling: an introduction and tutorial &\cite{ghaffarzadegan2024generative}\\
LLM-AIDSim: LLM-Enhanced Agent-Based Influence Diffusion Simulation in Social Networks &\cite{zhang2025llmaidsim}\\
Agent-Based Modelling Meets Generative AI in Social Network Simulations &\cite{ferraro2024agentbased}\\
Simulating Social Network with LLM Agents: An Analysis of Information Propagation and Echo Chambers & \cite{zheng2024socialnetwork}\\
\end{longtable}

\end{document}